\documentclass[preprint]{emulateapj}
\slugcomment{To appear in the Astronomical Journal}
\shortauthors{Zaritsky et al.}
\shorttitle{Condensed Baryon Fraction}

\begin{document}
\title{The Baryonic Tully-Fisher Relationship for S$^4$G Galaxies and the ``Condensed" Baryon Fraction of Galaxies} 
  
\author{Dennis Zaritsky\altaffilmark{1}, Helene Courtois\altaffilmark{2,3}, Juan-Carlos Mu\~noz-Mateos \altaffilmark{4}, Jenny Sorce\altaffilmark{2,5}, S. Erroz-Ferrer\altaffilmark{6,7}, S. {Comer\'{o}n}\altaffilmark{8,9},
D. A. Gadotti\altaffilmark{10}, A. Gil de Paz\altaffilmark{11}, J. L. Hinz\altaffilmark{12},
E. Laurikainen\altaffilmark{8,9},T. Kim\altaffilmark{4,10,13,14}, J. Laine\altaffilmark{8}, K. Men\'{e}ndez--Delmestre\altaffilmark{15}, T. Mizusawa\altaffilmark{4,16}, M. W. Regan\altaffilmark{17}, H. Salo\altaffilmark{8}, M. Seibert\altaffilmark{14}, K. Sheth\altaffilmark{4},  E. Athanassoula\altaffilmark{18}, A. Bosma\altaffilmark{18}, M. Cisternas\altaffilmark{6,7}, Luis C. Ho\altaffilmark{14,19}, and B. Holwerda\altaffilmark{20}}
\altaffiltext{1}{Steward Observatory, University of Arizona, 933 North Cherry Avenue, Tucson, AZ 85721, USA; dennis.zaritsky@gmail.com}
\altaffiltext{2}{Universit\'e Lyon 1, CNRS/IN2P3, Institut de Physique Nucl\'eaire, Lyon, France}
\altaffiltext{3}{Institute for Astronomy, University of Hawaii, 2680 Woodlawn Drive, Honolulu, HI 96822, USA}
\altaffiltext{4}{National Radio Astronomy Observatory/NAASC, 520 Edgemont Road, Charlottesville, VA 22903, USA}
\altaffiltext{5}{University of Potsdam: Leibniz-Institut f\"{u}r Astrophysik, Potsdam, Germany}
\altaffiltext{6}{Instituto de Astrof\'{i}sica de Canarias, V\'{i}a L\'{a}ctea s/n 38205 La Laguna, Spain}
\altaffiltext{7}{Departamento de Astrof\'{i}sica, Universidad de La Laguna, 38206 La Laguna, Spain}
\altaffiltext{8}{Astronomy Division, Department of Physical Sciences, FIN-90014 University of Oulu, P.O. Box 3000, Oulu, Finland}
\altaffiltext{9}{Finnish Centre of Astronomy with ESO (FINCA), University of Turku, V\"ais\"al\"antie 20, FI-21500, Piikki\"o, Finland}
\altaffiltext{10}{European Southern Observatory, Casilla 19001, Santiago 19, Chile}
\altaffiltext{11}{Departamento de Astrof\'{i}sica, Universidad Complutense de Madrid, 28040 Madrid, Spain}
\altaffiltext{12}{MMTO, University of Arizona, 933 North Cherry Avenue, Tucson, AZ 85721, USA}
\altaffiltext{13}{Astronomy Program, Department of Physics and Astronomy, Seoul National University, Seoul 151-742, Korea}
\altaffiltext{14}{The Observatories of the Carnegie Institution of Washington, 813 Santa Barbara Street, Pasadena, CA 91101, USA}
\altaffiltext{15}{Universidade Federal do Rio de Janeiro, Observat{\'o}rio do Valongo, Ladeira Pedro Ant{\^{o}}nio, 43, CEP 20080-090, Rio de Janeiro, Brazil}
\altaffiltext{16}{Florida Institute of Technology, Melbourne, FL 32901}
\altaffiltext{17}{Space Telescope Science Institute, 3700 San Martin Drive, Baltimore, MD 21218, USA}
\altaffiltext{18}{Aix Marseille Universit\'e, CNRS, LAM (Laboratoire d'Astrophysique de
Marseille) UMR 7326, 13388, Marseille, France}
\altaffiltext{19}{Kavli Institute for Astronomy and Astrophysics, Peking University, Beijing 100871, China}
\altaffiltext{20}{European Space Agency Research Fellow (ESTEC), Keplerlaan, 1, 2200 AG Noordwijk, The Netherlands}

\begin{abstract} 
We combine data from the Spitzer Survey for Stellar Structure in Galaxies (S$^4$G), a recently calibrated empirical stellar mass estimator from Eskew et al., and an extensive database of H{\small I} spectral line profiles to examine the baryonic Tully-Fisher (BTF) relation. We find 1) that the BTF has lower scatter than the classic Tully-Fisher (TF) relation and is better described as a linear relationship, confirming similar previous results, 2) that the inclusion of a radial scale in the BTF decreases the scatter but only modestly, as seen previously for the TF relation,  and 3) that the slope of the BTF, which we find to be $3.5\pm0.2$ ($\Delta \log M_{baryon}/\Delta \log v_c$), implies that on average a nearly constant fraction ($\sim 0.4$) of all baryons expected to be in a halo are ``condensed" onto the central region of rotationally supported galaxies. The condensed baryon fraction, $M_{baryon}/M_{total}$, is, to our measurement precision, nearly independent of galaxy circular velocity (our sample spans circular velocities, $v_c$, between 60 and 250 km s$^{-1}$, but is extended to $v_c\sim 10$ km s$^{-1}$ using data from the literature). The observed galaxy-to-galaxy scatter in this fraction is generally $\le$ a factor of 2 despite fairly liberal selection criteria. These results imply that cooling and heating processes, such as cold vs. hot accretion, mass loss due to stellar winds, and AGN driven feedback, to the degree that they affect the global galactic properties involved in the BTF, are independent of halo mass for galaxies with $10 < v_c < 250$ km s$^{-1}$ and typically introduce no more than a factor of two range in the resulting $M_{baryon}/M_{total}$. Recent simulations by Aumer et al. of a small sample of disk galaxies are in excellent agreement with our data, suggesting that current simulations are capable of reproducing the global properties of individual disk galaxies. More detailed comparison to models using the BTF holds great promise, but awaits improved determinations of the stellar masses.
\end{abstract}


\section{Introduction}
\label{sec:intro}

Empirical galaxy scaling relations are a testament to the existence of underlying physical principles of galaxy formation. Among such scaling relations, the relationship between the width of the neutral hydrogen line and the luminosity of a galaxy \citep{tf} stands as one of the most useful \citep[for example see its use as a distance estimator;][]{freedman} and constraining \citep[for example see its use to test complicated baryonic physics in dark matter halos;][]{steinmetz}. 

Even though the Tully-Fisher relationship (hereafter the TF relation) has been extensively vetted and explored \citep{freedman10}, some questions remained unanswered. Foremost are those relating to the physical origin of the relationship. Although vaguely related to the Virial theorem, the relationship is not simply a recasting of that theorem \citep[cf.][]{mcgaugh98,mcgaugh00,z12a}. In particular, one can imagine constructing two galaxies with the same rotation curve, but extending the stellar disk a few times farther out in one. Although both galaxies would satisfy the Virial theorem, the two galaxies could not lie on the same relationship between rotation velocity and luminosity. Nature, apparently, cannot envision two such galaxies. 

However, the TF relation in its simplest incarnation is not a complete description of all disk galaxies. It fails to match the characteristics of some faint, gas rich galaxies \citep{carignan,persic,meurer,mcgaugh00}. A scaling relationship is recovered if one recasts it as one between rotation velocity and baryonic mass rather than just luminosity \citep{freeman,walker,mcgaugh00,verheijen,geha}, and this is referred to as the baryonic TF (hereafter, BTF).  These results suggest that the original TF exists because for most galaxies in TF studies the luminosity is a reasonably precise proxy for stellar mass and the gas mass is negligible. 

In a quest to uncover more clues and improve the fidelity of the TF as a distance estimator, investigators have long sought additional parameters that would help reduce the scatter in the TF relation --- a search for a second, or even third, parameter beyond the rotational velocity \citep[see, for examples,][]{strauss,zwaan,sprayberry,courteau97,courteau,mcgaugh05b,kassin,hall}. In this context, evidence for the importance of a scaling radius has been presented \citep[cf.][]{kassin,hall}, although many other studies found no such dependence \citep{zwaan, sprayberry, courteau, mcgaugh05b}. 

The principal scaling relationship for early type galaxies, the Fundamental Plane \citep{dd87,d87}, includes a kinematic term, a scaling radius, and a surface brightness term. Following along that line of reasoning, we have explored a joint scaling relationship for both late and early types that has the flavor of the Fundamental Plane  \citep{z06a,z06b,z08,z12a}. Given its antecedent and geometry, it is referred to as the Fundamental Manifold. The implication of that work in the current context is that disk galaxies should show residuals from the standard TF that correlate with the half light radius. In this study, we examine whether the introduction of this scaling radius improves the BTF as well.

Finally, the BTF has great promise to provide detailed, quantitative tests of galaxy formation and evolution models \citep[for some examples of this approach see:][]{mayer,gnedin,governato,avila,gurovich,mcgaugh12,aumer}. The limiting uncertainty in such tests has been the stellar mass determination. While prescriptions for the mass-to-light ratio based on observed colors are available \citep{bell}, these depend sensitively, at a level of precision that incapacitates the envisioned test, on the adopted stellar initial mass function and star formation history. Because these techniques are based on stellar population modeling they also carry fundamental uncertainties on the modeling of rare, but luminous, phases of stellar evolution \citep{maraston,conroy}.  As an alternative, \cite{eskew} used the resolved stellar population study of the Large Magellanic Cloud \citep{harris} and {\sl Spitzer} images \citep{meixner} to calibrate the conversion of 3.6 and 4.5 $\mu$m luminosities to stellar mass. This approach bypasses some, but not all, of the weaknesses mentioned above \citep[see][for a discussion of the relative merits]{eskew}.  The use of the IR tracers also mitigates the role of internal extinction and detailed work at these wavelengths aims to correct for dust emission as well \citep{meidt}. With the advent of this new stellar mass estimator and large samples of galaxies observed with the {\sl Spitzer} Space Telescope, we now return to reexamine the BTF.

In this study we combine homogeneous, high quality IR data (3.6$\mu$m and 4.5$\mu$m surface photometry) from the {\sl Spitzer} Survey for Stellar Structure in Galaxies \citep[S$^4$G;][]{sheth}, which is less susceptible to extinction than optical data and has a more uniform stellar mass-to-light ratio, the \cite{eskew} stellar mass estimator, and  H{\small I} spectral line profiles from the literature curated by the Cosmic Flows project \citep{courtois11} to re-explore some of the questions raised by both the standard and baryonic Tully-Fisher relationship. A strength of this study is that the parent sample is fairly broadly selected to be a magnitude limited, volume limited sample. As such, the results are representative of galaxies in general rather than of a ``pristine" sample intended to provide the tightest scaling relation or most reliable distances. In \S2 we describe the data, present our findings regarding the BTF and the role of other parameters in \S3, and conclude in \S4. 

\section{The Data and Measurements}
\label{sec:data}

The photometric data, from which we obtain the measurements of the infrared luminosity, half light radii, and inclinations come from the S$^4$G dataset \citep{sheth} and subsequent analysis described by \cite{munoz} and \cite{salo}. The basic data consist of images obtained with the {\sl Spitzer} Space Telescope \citep{werner} with the IRAC instrument \citep{irac} during its warm mission, so limited to the 3.6$\mu$m and 4.5$\mu$m
channels, of 2352 galaxies in the local universe. The data processing, masking, and photometry are described by \cite{munoz}, with additional model fitting described by \cite{salo}.  The data have been used in a variety of studies that can ultimately complement that of scaling relations, including those of \cite{buta}, \cite{comeron}, \cite{holwerda}, and \cite{z13}. 

\begin{deluxetable*}{rrrrrrrrrr}
%\rotate
\tablewidth{0pt}
\tablecaption{Primary Sample}
\tablehead{
\colhead{PGC Number}&\colhead{Alternative Name}&\colhead{$\log v_c$}&\colhead{$\log M_{baryon}$}&\colhead{$\log M_*$}&\colhead{$\log M_{atomic}$}&\colhead{$\log M_{mol}$}&\colhead{$i$}&\colhead{T-Type}&\colhead{$cz$}\\
&&\colhead{[km s$^{-1}$]}&\colhead{[$M_\odot$]}&\colhead{[$M_\odot$]}&\colhead{[$M_\odot$]}&\colhead{[$M_\odot$]}&\colhead{[$^\circ$]}&&\colhead{[km s$^{-1}$]}\\
}
\startdata
72957&ESO012-010&2.07&10.01&9.48&9.81&8.77&62&7.7&1925\\
181&ESO012-014&1.72&9.87&9.17&9.78&$-$9.99&65&9.0&1936\\
13695&ESO015-001&1.82&9.72&8.96&9.63&$-$9.99&65&9.6&1659\\
13931&ESO054-021&2.04&10.13&9.73&9.87&8.68&60&7.9&1424\\
 2445&ESO079-005&1.96&9.82&9.28&9.59&8.90&53&7.0&1599\\
16299&ESO085-014&1.97&9.86&9.42&9.67&$-$9.99&68&9.0&1420\\
16780&ESO085-047&1.51&9.42&8.74&9.32&$-$9.99&57&9.0&1491\\
18051&ESO120-021&1.81&9.32&8.43&9.26&$-$9.99&60&10.0&1364\\
\enddata
\break
\label{tab:data}
\tablenote{This table is available in its entirety in a machine-readable form in the online journal. A portion is shown here for guidance regarding its form and content.}
\end{deluxetable*}

Since 2009, the Cosmic Flows project (CF) has gathered all the digital H{\small I} spectra available from the public archives of the largest radio-telescopes worldwide and re-measured them in a consistent way. Two sub-projects of CF, at Green Bank in the USA and at Parkes in Australia \citep{courtois11}, complete the archives for targets without previous observations that are adequate for TF studies.

The main goal of CF is to map the all-sky peculiar velocity field at redshift zero and reconstruct the underlying dark matter distribution. For that purpose, tens of thousand of galaxy line widths were measured with a new robust method described by \citet{courtois09,courtois11}. Briefly, the line width parameter, $W_{m50}$, is a measure of the H{\small I} profile width at $50\%$ of a specially calculated estimate of the maximum flux within the velocity range encompassing $90\%$ of the total H{\small I} flux (details provided in the cited references and computer code available from H. Courtois). This measurement is then transformed into the parameter $W_{mx}^{av}$ by correcting for the slight relativistic broadening,  broadening due to finite spectral resolution, internal turbulent motions \cite[see Eq.2 in][]{courtois11}, and averaged if there are multiple good measurements. Further details can be found in \citet{courtois09,courtois11} and \cite{tully12}. The question of whether one should use a direct measurement of the maximum of the rotation curve, the width of the H{\small I} profile as a proxy for that maximum velocity, or the rotational velocity measured over the flattest part of the rotation curve is a long-standing one in TF work that we do not have the data to address. Given the data available to us, we use the width of the H{\small I} profile, and in particular utilize $W_{m50}$, which has been shown to be superior to alternative parameterizations of the line width \citep{courtois09}. 

\begin{figure*}
\plotone{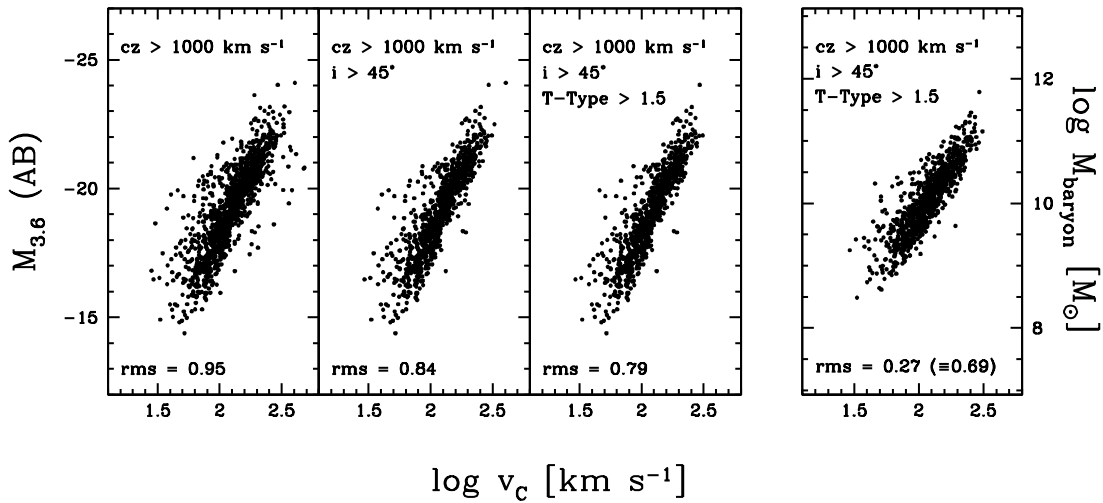}
\vskip -2in
\caption{Scatter and the IR Tully-Fisher Relation. We begin with the TF relation, with only a basic redshift cut in the leftmost panel to mitigate large distance errors introduced by peculiar motions. The rms scatter about a linear least square fit is presented as a quality fiducial. In the subsequent panel, we apply an inclination cut to eliminate galaxies that are sufficiently face-on that corrections to the rotation velocity become large. This cut removes a number of outliers and reduces the rms deviation about a new best fit line. In the third panel we plot the TF after removing early type galaxies, which may be significantly supported by velocity dispersions rather than rotation. Again, some improvement is seen. This plot includes what we call our primary sample (see Table \ref{tab:criteria}). In the final panel, we plot the BTF for our primary sample, calculated as explained in the text, and the equivalent rms (correcting for the factor of 2.5 in magnitudes) is now 0.69, significantly smaller than that for the TF of the same sample.}
\label{fig:TF}
\end{figure*}

We currently have coherent H{\small I} measurements for 16,121 galaxies. Of those, 12,189 are of sufficient quality for distance measurements with TF.
This catalog is available for public use at the Extragalactic Distance Database (EDD) website\footnote{http://edd.ifa.hawaii.edu; catalog ``All Digital H{\small I}"} and we call it the ``All Digital H{\small I} catalog".
Several other parameters available are included \citep{tully09} such as the integrated H{\small I} line fluxes computed from the H{\small I} lines, which have a flux calibration uncertainty of about 10 to 15\%, and the average heliocentric velocities.

To calculate the TF parameters, we define our maximum circulate velocity, $v_c$, as $v_c=W_{mx}^{av}/(2\sin i)$, where $i$ is the inclination. 
We adopt inclinations from our analysis of the S$^4$G images \citep{salo}, 
which includes multi-component galaxy modeling and produces an inclination measurement for the disk component. We find that the TF residuals using this inclination are smaller than those using B-band inclinations from RC3 \citep{rc3}, a photometric band that emphasizes the disk over the bulge. Nevertheless, this approach, like any that use photometric measurements to estimate inclination and gaseous kinematics to measure the rotation velocities, implicitly assumes that the stars and gas are coplanar and that the stellar isophotes are intrinsically round.

One can then derive a distance using 

$$\log D_L =(m_{3.6}+20.34+9.74(\log v_c-2.5)-25)/5,$$

\noindent
which comes from a recent IR TF calibration \citep{sorce13}, where $m_{3.6}$ is the total apparent magnitude derived in the {\sl Spitzer} photometric band at 3.6 $\mu$m. However, here we use the smooth Hubble flow distances, calculated from the CMB-frame recessional velocities, to avoid circularity when examining the TF residuals. We adopt H$_0$ = 70 km s$^{-1}$ Mpc$^{-1}$, $\Omega_m = 0.3$, and $\Omega_\Lambda = 0.7$ and account for the differences between luminosity and angular distances, the latter used for our calculation of radii.
Finally, for our calculation of the BTF we use the H{\small I} flux parameter to compute the H{\small I} mass of a given galaxy using the equation

$$M_{HI} =2.36 \times 100000 \times D_L^2 \times F$$

\noindent
in units of $10^5 M_\odot$, where $M_{HI}$ is the H{\small I} gas mass, $D_L$ is the luminosity distance in Mpc, and $F$ is the flux integrated within the H{\small I} line profile in units of Jy km s$^{-1}$.
Together, we have 1468 galaxies for which we have the necessary S$^4$G and H{\small I} data.

\section{Results}

\subsection{The S$^4$G TF Relation and Sample Selection}

The basic TF relation for S$^4$G galaxies is shown in the leftmost panel of Figure \ref{fig:TF}, with the only exclusion being galaxies that have recessional velocities that are less than 1000 km s$^{-1}$.  This criterion removes 218 galaxies from consideration. The recessional velocity cut is the lowest possible to avoid including galaxies whose peculiar velocities, which can range as high as a few hundred km s$^{-1}$ except in the centers of galaxy clusters where they can be larger, introduce significant errors in the inferred distances. We will show later that this cut needs to be significantly increased. 

\begin{deluxetable}{lrr}
\tablewidth{0pt}
\tablecaption{Sample Selection Criteria}
\tablehead{
&\colhead{Primary Sample}&\colhead{Low Scatter Subsample}\\
}
\startdata
extant H{\small I} observation&yes&yes\\
recessional velocity&$cz > 1000$ km s$^{-1}$&$ cz > 2000$ km s$^{-1}$\\
inclination&$45^\circ< i$&$45^\circ < i < 80^\circ$\\
T-Type&$1.5 < T $&$3 < T < 8$\\
\enddata
\label{tab:criteria}
\end{deluxetable}

The goal of this study is not to present the functional form of an optimal IR TF relation (see \cite{sorce13} for that line of inquiry), but we do want to compare the scatter for different incarnations of the scaling relations. Therefore, we fit a line using least squares to provide a fiducial against which to compare the scatter and quote the rms about that fit in the lower left of the panel. 
The existence of a basic TF relation is evident, although there is obvious scatter. Whether this scatter is reducible either through the exclusion of galaxies that are ill-suited for this measurement, or the inclusion of, or correction for, another physical parameter in the scaling relation, or whether it is irreducible observational noise is what we now explore.

We have already made one defensible selection cut on recessional velocity, we next explore one on disk inclination relative to the line of sight. Because the measured rotational velocity must be corrected for inclination, as the disks become more face on this correction becomes larger and uncertainties in the inclination measurement dominate. Given the large size of the sample, we can afford to be conservative and ignore systems that are relatively face-on, thereby avoiding this problem entirely. We have selected to include only those galaxies with inclinations $> 45^\circ$. We will return to justify this specific selection below (\S3.2). This cut removes an additional 278 galaxies from consideration. Optical studies also face problems at large inclination because of the required high extinction corrections to the total magnitude. By observing at  3.6$\mu$m we find that this problem is significantly reduced, as we will show later (\S3.2). The result of our lower bound on inclination is seen in the second panel in Figure \ref{fig:TF}. A number of outliers are removed and the rms scatter is reduced. 

Next, we consider that the standard TF relation does not describe early-type galaxies, which can have stellar components with almost no rotation that are quite luminous. To eliminate these galaxies, we consider only galaxies of T-Type $>$ 1.5 (Sab's and later). Again, this helps reduce the scatter as can be seen in the third panel of Figure \ref{fig:TF} and reduces the sample by a further 79 galaxies to a total of 903 galaxies. This set of galaxies is what we refer to as our primary sample.

Despite the improvements that these criteria have realized, there are at least two problems they have not fully addressed. First, the scatter upward from the ridge line remains significant. It is curious, and perhaps telling, that the lower edge of the ridgeline is particularly sharp and well-defined, suggesting that random observational errors are not the dominant cause of the remaining scatter. Second, the relationship is not a straight line but rather has a downward kink at $\log v_c \sim 2$. This is exactly the class of feature that has been seen in studies advocating the BTF over the classic TF, particularly for low-luminosity galaxies \citep{mcgaugh00}. 

\subsection{The Baryonic TF Relation}

Following that line of reasoning, we now calculate the baryonic mass of these galaxies. The stellar mass we obtain using the IR calibration of \cite{eskew} based on spatially resolved stellar population studies of the Large Magellanic Cloud \citep{harris,meixner} that enables us to convert the combination of 3.6 and 4.5$\mu$m fluxes to a corresponding number of solar masses ($M_*  = 10^{5.65} F_{36}^{2.85} F_{45}^{-1.85}$). This estimator has subsequently been confirmed both with comparison to SDSS-derived stellar masses \citep{cybulski} and with detailed dust/star decompositions of the IR flux \citep{que}. To estimate the gas mass we use the relationship described above to obtain $M_{HI}$, adopt a sliding scale in the correction for the $H_2$ mass as a function of galaxy type \citep{knezek,young}, and correct for the mass in He and metals by multiplying by 1.4. 
We adopt the parameterization of the dependence of molecular mass fraction with galaxy type presented by \cite{mcg}: $M_{H_2}/M_{HI} = 3.7 - 0.8T + 0.043T^2$, where $T$ is the galaxy T-type. In a small number of cases this formula results in an unphysical value (negative, but small) for the molecular mass and in those cases we set it to 0. 
The correction for the molecular gas mass is often ignored because it is negligible in the relevant class of galaxy \cite[see][for one such example]{geha}. We do include it, but find that it makes only a modest difference for our sample.
Alternative prescriptions, based for example on the surface density of H{\small I} also exist \citep{leroy,mcgaugh12}, but we do not have the necessary information to apply this approach and the estimation of the molecular mass is a minor source of uncertainty here. The resulting mass estimates, and other parameters necessary for the BTF, are presented in Table \ref{tab:data} for our primary sample.

We refer to the sum of the stellar and gaseous mass calculated in this manner as the baryonic mass, $M_{baryon}$, and plot it as a function of $v_c$ in the rightmost panel of Figure 1. It is important to note, however, that despite the common usage of the term baryonic mass in this context we have not included the potential contributions of either extremely cold material (few K) that evades the molecular measurements \citep[see, for examples,][and references therein]{pfenniger} or warm ($>$ 10$^5$ K) gas that may be present. Furthermore, we have not included the presence of baryonic material at large radii. Evidence for extended distributions of stars has been presented in early-types \citep{tal}, for extended star formation as a general feature in a significant fraction of late-types \citep{thilker, gdp, zc, hf}, and for dust at large radii \citep{z94,nelson,menard}. We do not, therefore, expect this to be the full baryon accounting of galaxies. Instead, this represents the fraction of baryons that have ``condensed" onto the central parts of galaxies. 
\begin{figure}
\plotone{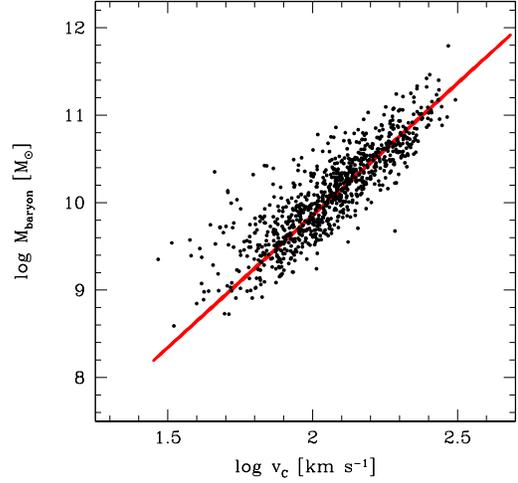}
\caption{The BTF for our primary sample (see Table \ref{tab:criteria}). The line represents our fiducial about which we calculate residuals, derived using a bisector regression fit \citep{isobe}. The fitted line for galaxies with $\log v_c > 1.7$ has slope $3.022 \pm 0.007$, but the internal uncertainty is a gross underestimate of the true uncertainty.}
\label{fig:btf}
\end{figure}

Examining the BTF, we find that once again the scatter has improved (to an equivalent scatter of 0.69 after correcting for the factor of 2.5 present in the magnitudes that is not present in the log(masses)) and that this improvement comes not from removing outliers but from straightening the ``kink" seen in the previous panels at log $v_c \sim 2$. It is evident that the BTF, with nearby, face-on, and early-type galaxies removed, is the tightest scaling relation of those examined so far. For distance work, we recommend using the BTF if possible, particularly if the sample includes galaxies with $v_c <125$ km s$^{-1}$. However, the classic TF can show remarkably small scatter in appropriately selected samples. \cite{verheijen} finds that in a sample of galaxies in the Ursa Major cluster there is no evidence for any intrinsic scatter. For some lines of inquiry, a well-selected sample and exquisite data provide unique insights. Here, however, we are considering the broader disk galaxy population, warts and all.

Despite the significant improvements obtained in the scaling relations described above, the upward scatter of points remains (Figure \ref{fig:btf}). The asymmetric nature of that scatter again suggests that random observational errors are not solely at work. Historically, studies of TF and other scaling relations have made a number of difficult-to-reproduce selection cuts to clean the samples of such outliers, for example often based on a galaxy's visual appearance or on the nature of the H{\small I} spectral line profile. Such culling may indeed be appropriate for certain work, such as a determination of H$_0$, but are more questionable  for comparison to models of galaxy evolution where the whole range of the population needs to be explored. This line of reasoning is particularly important when comparing to simulations, where it is difficult to control for different influences particularly because one does not know which ones are critical. For example, while requiring that a galaxy not suffer a major merger for $z < 1$ may ensure that the simulated galaxy contains a thin disk there is no guarantee that it will match specific visual criteria for interaction signatures or deviations from a classic double-horn H{\small I} spectral line profile. Current, high resolution simulations \citep[eg.,][]{aumer} are limited in the number of galaxies simulated, so they do attempt to target a constrained class of galaxy, such as those with a thin disk, but the resulting relation to empirical criteria is uncertain. On the observational side there is, therefore, a delicate balance in determining which selection criteria are well motivated in an effort to recover the intrinsic scatter in a scaling relation and which may artificially reduce the scatter. We now discuss our approach at refining  appropriate selection criteria or parameters in an effort to uncover a realistic scaling relation with the minimal scatter that outlines the true underlying relationship.

\begin{figure}
\plotone{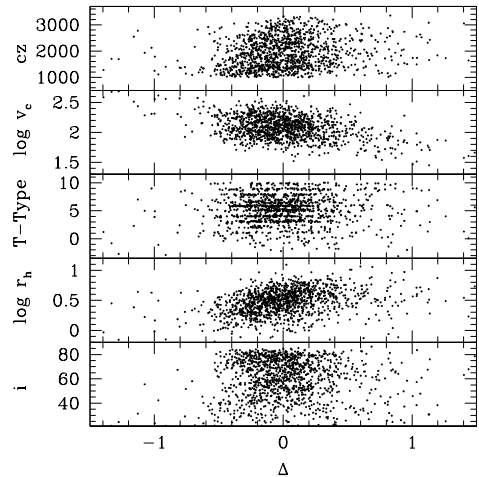}
\caption{Residuals from the BTF and various galaxy parameters. Here we plot our complete sample of 1478 galaxies (a few lie off the plot boundaries). The units on inclination, $i$, are degrees, on $r_h$ kpc, and on $v_c$ and $cz$ km sec$^{-1}$.}
\label{fig:resids}
\end{figure}

Beginning with the BTF relation, reproduced in Figure \ref{fig:btf}, we examine the nature of the residuals about a fitted line (slope = $3.022 \pm 0.007$, for galaxies with $\log v_c > 1.7$ to eliminate the tail of outliers). The fit is the result of a bisector regression fit \citep{isobe} and we will return to a discussion of choice of fitting method below because indeed the fitting algorithm can be a significant source of uncertainty in this type of work.  Not only for this reason, the internal uncertainty is a gross underestimate of the true uncertainty. For now this line serves as a fiducial against which to calculate residuals. We plot the residuals about the line relative to various characteristics of the galaxies in Figure \ref{fig:resids}. We are in general looking for two broad classes of phenomena. First, we look for a significant increase in scatter over a limited parameter range. Such a feature suggests either that we cannot measure galaxy properties sufficiently well for this parameter range, for example at low inclinations, or that the TF relationship is not applicable, for example for early type galaxies. We find both of these effects in Figure \ref{fig:resids}. Examining the panel showing the residuals, $\Delta$, relative to inclination, there is no concentration of galaxies about $\Delta = 0$ for inclinations $<$ 40$^\circ$. This justifies our cut at $45^\circ$. Examining the panel showing the relation between $\Delta$ and T-Type, there is no concentration toward $\Delta = 0$ for T-Type $<$ 2, supporting our criterion of T-Type $> 1.5$.  Second, we look for a systematic trend in residuals suggesting either a systematic error or additional physical information. There are two fairly evident such trends in Figure \ref{fig:resids}. First, there is that of $\Delta$ with $v_c$, although that is mostly defined by long tails of high residuals. In the core of the distribution such a correlation is not so clear. Second, there is a trend of $\Delta$ with the half light radius, $r_h$, that is mostly present in the core of the distribution, but not in the tail of high residuals. We will examine these in greater detail, once we have dealt with additional sources of scatter that we identify on the basis of this Figure.

Two sources of scatter, low inclinations and early T-Types, we have already addressed. We identify two additional sources of scatter that we have not yet mitigated. First, looking at the panel showing $\Delta$ vs. $v_c$, and in particular the tail of large residuals, we see significantly larger scatter among galaxies with lower recessional velocities. We suspect that this result is caused by larger relative distance errors due to peculiar velocities for these less luminous galaxies, which in a flux limited sample tend to be nearer. On the other hand, without additional information, we cannot exclude that the larger scatter is due to greater intrinsic scatter among fainter galaxies. However, previous studies that focused on low luminosity galaxies found that they do follow the BTF with low scatter \citep{geha, mcgaugh12}, so we conclude that peculiar velocities are the likely culprit. We therefore raise our cut to include only galaxies with $cz > 2000$ km sec$^{-1}$ for a new ``low scatter" subsample. Second, examining the panel that contains the T-Types, we see that there is no concentration about $\Delta = 0$ for the largest T-types. This can either be interpreted as a true physical failing of the scaling relation for these galaxies or as evidence that either the rotation velocities or inclinations have much larger uncertainties for these galaxies. Regardless of the origin of the scatter, we will exclude galaxies with T-Type $\ge 8$ from our ``low scatter" subsample. Lastly, there appears to be a slight offset in the mean residual for galaxies with inclinations close to 90$^\circ$, perhaps as a result of high internal extinction. To be conservative, we also increase our lower bound on T-Type to 3 and impose an upper inclination bound of 80$^\circ$ for our low scatter subsample. The primary and low-scatter subsample criteria are reprised in Table \ref{tab:criteria}.

\begin{figure}
\plotone{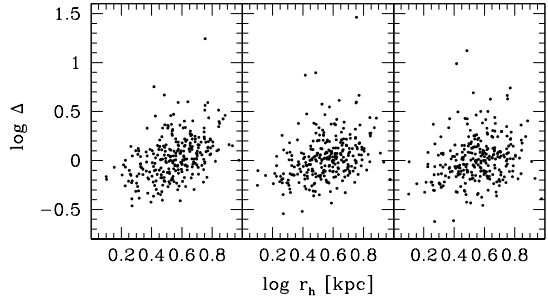}
\caption{Residuals from the BTF and $r_h$ for our low scatter subsample (Table \ref{tab:criteria}). We show the dependence of the residuals in the BTF with $r_h$ for three different choices of BTF slope. The central panel shows the residuals about a slope = 3 line, while the left and right show the residuals about lines with slopes 2.5 and 3.5 respectively. The choice of slope does affect the distribution of residuals, but even in the right panel, where the correlation between residual and $r_h$ is weakest, a Spearman rank correlation analysis indicates that there is a $1.5\times 10^{-5}$ chance that these two quantities are unrelated.}
\label{fig:re}
\end{figure} 

\begin{figure}
\plotone{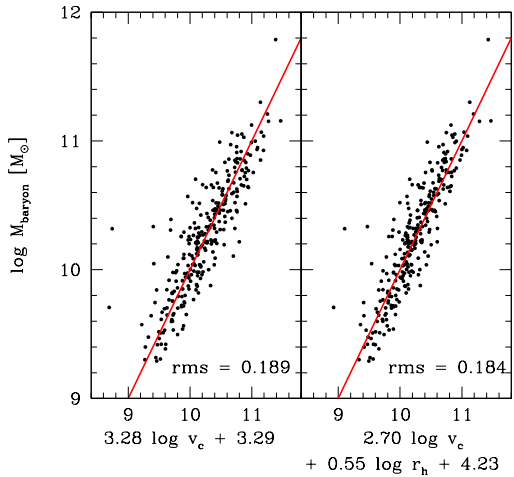}
\caption{Comparison of fit vs. measured $M_{baryon}$ for BTF and scale-dependent BTF for our low scatter subsample. The plotted lines are the 1:1 relation. Improvement provided by including scale-dependence is noticeable but modest, and certainly not the primary source of the scatter.}
\label{fig:fit1}
\end{figure} 

Implementing these additional cuts, we now return to the systematic behavior of the residuals with respect to $r_h$. In Figure \ref{fig:re} we show this behavior relative to the slope = 3 line shown in Figure \ref{fig:btf}, and for slope 2.5 and 3.5 lines. 
Regardless of the fiducial against which we calculate the residuals, there is a relationship between that residual and $r_h$. Even for what is visually the weakest case, that using the slope = 3.5 line (rightmost panel), a Spearman rank correlation coefficient concludes that the chance the points are randomly distributed with respect to $r_h$ is $<2 \times10^{-5}$. We conclude that the scatter in the BTF will be reduced by introducing a dependence on $r_h$, and we return to this below. The new cuts also greatly reduce the high and low $\Delta$ tails in the $\log v_c - \Delta$ distribution. 

Using the low-scatter subsample, we now examine the BTF and the effect of the inclusion of an $r_h$ term in Figure \ref{fig:fit1}.
Fitting a linear relationship between $\log M_{baryon}$ and $\log v_c$, using here an ordinary least-squares approach and rejecting galaxies that have residuals $> 0.6$ dex, results in a slope of 3.28 and an rms about the fit of 0.189. To include an $r_h$ dependence, we fit an equation of the form $\log M_{baryon} = A \log v_c + B \log r_h + C$, again excluding galaxies with residuals about the fit $> 0.6$.  The rms is only slightly lower at 0.184. The gains achieved by including $r_h$ in the fit are detectable, but indeed modest. 
It is quite likely that the results of this analysis depend critically on the radius at which one measures the rotational velocity because of the sensitivity of the peak of the rotation curve on the degree of mass concentration and the lack of such sensitivity in the asymptotic value of the rotation curve. As such, our results apply to the use of W$_{50}$ and are likely not directly applicable to other measurements of $v_c$ \citep[see for example,][]{verheijen}. 
We conclude that for disk galaxies over the present parameter range there is no pressing requirement to include an $r_h$ scaling and so continue our discussion with the standard BTF.

The scatter remains asymmetric, although there are now only a few outlying galaxies. Because we have removed the nearby galaxies that were susceptible to large distance errors,  these outliers are probably due to dynamical effects, such as interactions, that disturb the H{\small I} velocity field \citep{ho}. Similar tails to the distribution have been noticed before \citep{ho} and are not a unique artifact of our data. Indeed careful pruning of samples based on interaction signatures, either in the morphology or the nature of the H{\small I} spectral line profile often removes such outliers \citep{verheijen}. This population, if indeed the deviations are physically driven, should be included in comparisons to simulations because it is difficult to ensure that attempts to prune both the theoretical and empirical samples of ``disturbed" galaxies will fairly and completely reproduce the selection.

\begin{figure*}
\plotone{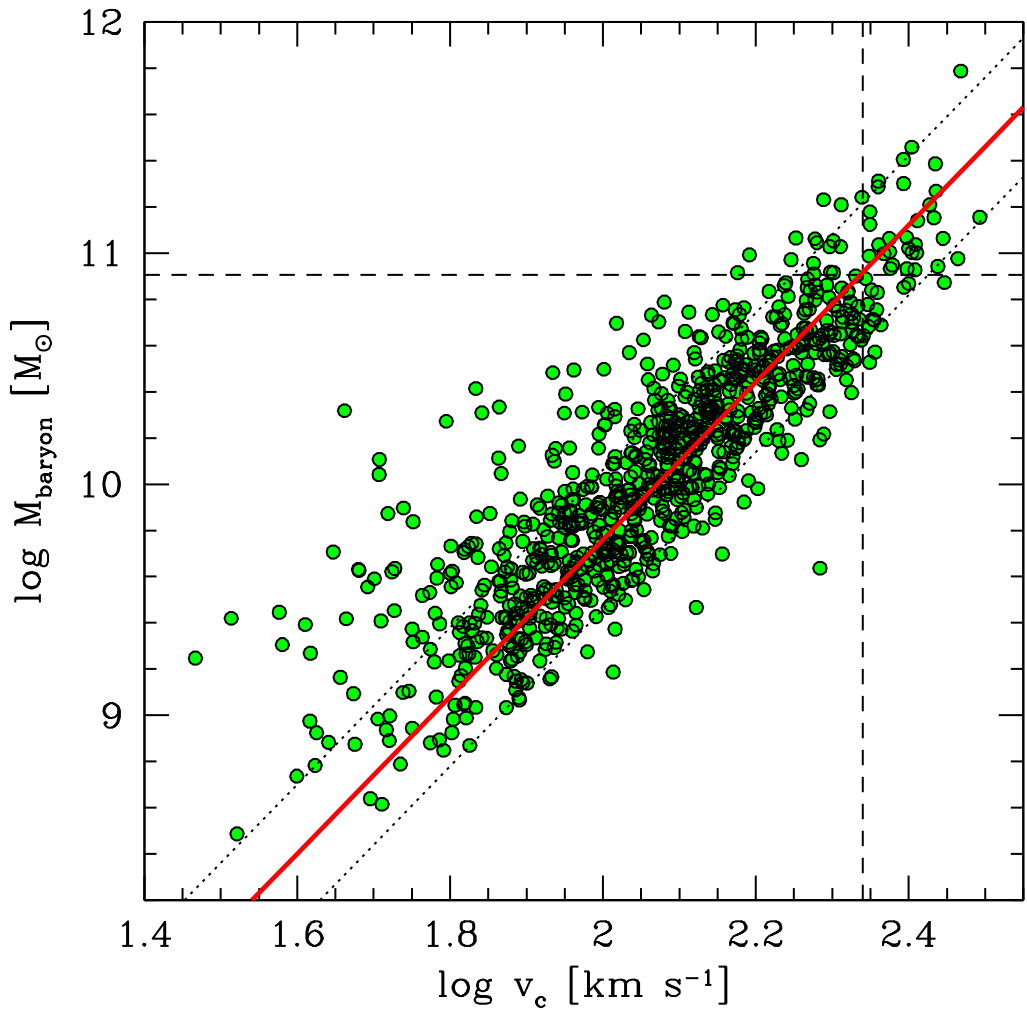}
\caption{ Baryonic Tully Fisher compared to a simple model in which galaxies condense out  42\% of their baryons as stars or cold gas (thick red line) and follow the halo mass$- v_c$ relation from \cite{bullock}. The dashed lines show the position of the MW that is used to provide a coarse normalization of the relationship. The inclined dotted lines show factors of two deviation from the model in the condensed baryon fraction. The simple description of a constant condensed baryon fraction is an excellent description of the sample mean, with variations of up to a factor two for individual galaxies allowed by the current level of scatter.}
\label{fig:bf}
\end{figure*}

\subsection{The ``Condensed" Baryon Content of Galaxies}

As shown by our consideration of the $r_h$ term, the dominant term in the 
mass estimation is $v_c$.  Therefore, we return to consider the BTF, reprised in Figure \ref{fig:bf}
using our primary sample. In the simplest model of galaxies we envision, the mass of a halo of characteristic velocity $v_c$ is proportional to $v_c^3$. Two powers of $v_c$ come from $M \propto v^2r$ and the third comes from the virial radius, $r$, having the characteristic that $r \propto v_c$, which is generically expected for dark matter halos \citep[see][for a review of this topic]{mo}. If the fraction of baryons that settle, or condense, onto the central portion of the galaxy is a fixed number, then in this simple model we would expect a slope = 3 line to fit the data as plotted. The normalization of that line would then provide the numerical value of that fixed fraction.
In detail, when the growth of halos is simulated in a cosmological context the actual value of how halo mass depends on circular velocity differs slightly from our naive expectation \citep[slope = 3.4;][]{bullock}. 

To compare the theoretical expectation to our data, we superpose the expectation on the data in Figure \ref{fig:bf}. Using the Milky Way (MW) as a reference, adopting $v_c \sim$ 220 km sec$^{-1}$ and total mass of $\sim 1.2 \times 10^{12} M_\odot$, an estimate that has remained relatively unchanged since observations of distant test particles became available \citep{z89,watkins, mbk, barber}, we find that a galaxy with the total mass of the MW would have a baryonic mass, defined in the manner we have done so far in terms of the stellar mass plus 1.4 times the H{\small I} mass plus the correction for molecular gas, of 8.3$\times 10^{10} M_\odot$. When compared with the total mass, we derive a ``condensed" baryon mass fraction ($M_{baryon}/M_{total}$) of 0.07. If we adopt that $M_{total} \propto v_c^{3.4}$, following \cite{bullock}, and posit that on average galaxies have the same condensed matter fraction, then we arrive at the solid red line in the Figure. 
The concurrence, in the mean, is manifestly excellent.

The condensed baryon fraction we derive, 0.07, should be compared with the universal value of the baryon to total matter ratio \citep[0.1649 from WMAP9;][]{wmap}, which then suggests that rotationally supported galaxies consistently condense out, either as stars or cold gas, about 40\% of all of their baryons onto their central, luminous, regions regardless of their circular velocity, over the range of circular velocities shown in the Figure. The $v_c$ range can be extended to lower values of $v_c$ with observations that have targeted low luminosity galaxies \citep{mcgaugh98,geha,trach,stark} and the relationship still holds (see below).
In other words, we find no evidence of significant differential baryonic mass loss within this sample, confirming previous work on the BTF \citep{mcgaugh05a,geha,trach,mcgaugh12}. The apparent consistency in this baryonic fraction is once again a testament to the regularity underlying galaxy formation and evolution, but is in tension with other estimates of the baryonic fraction (see end of \S3.4).

The degree of scatter observed is significant, the dotted lines in Figure \ref{fig:bf} represent a factor of two difference in either direction. The bulk of the galaxies lie within this factor of two range. The majority of the outliers well above this are removed once the cuts implemented for our low-scatter subsample are applied, and we explore the BTF for that sample below. The interesting question is how much of the scatter within the factor of two level is intrinsic. As we mentioned previously, some TF studies \citep{verheijen} have found no evidence for intrinsic scatter. Our data are not suited to such an exploration because we do not have spatially resolved rotation curves, but we again stress that in terms of comparison to simulations a ``pure" galaxy sample may not be the most appropriate comparison sample.

\subsection{Comparison to Previous Studies and Simulations}

The results on the condensed fraction and comparison to simulations are potentially highly constraining. \cite{mcgaugh12} explore in detail comparisons to different theoretical predictions, both for $\Lambda$CDM and non-standard (MOND) models. However, as he notes, a fundamental limitation of the current BTF is the determination of the stellar masses. In particular, the slope of the BTF can range between 3 and 4 depending on which photometric bands are used and significant variations are also possible depending on the fitting algorithm and treatment of outliers. Indeed, previous studies \citep{avila,hall} have found shallower BTF slopes than that presented by \cite{mcgaugh12}, 4,  and our use of a different stellar mass estimator
results in data that are consistent with a BTF of slope 3.4 (Figure \ref{fig:bf}).

\cite{stark} noted that one can avoid the uncertainties in the stellar modeling by studying galaxies in which the baryonic mass is dominated by the gas. Although determining the gas mass has its own issues (unknown molecular gas mass, corrections for metallicity), this approach provides a check on the results that are based on systems where stellar masses dominate. It does, however, assume that a single linear BTF applies to all rotationally supported galaxies. We exploit this idea by comparing the results for gas-dominated galaxies directly to those for stellar-dominated galaxies. By selecting only those galaxies in which one or the other component dominates, we separate the relative mass normalizations. We adopt the \cite{mcgaugh12} measurements for galaxies in which the stellar mass contributes $<$ 20\% of the baryonic mass (these are typically low $v_c$ galaxies that are not in the S$^4$G sample) and compare to those galaxies in our ``low scatter" subsample that have gas masses that account for $<$ 20\% of the total baryonic mass.  

\begin{figure*}
\plotone{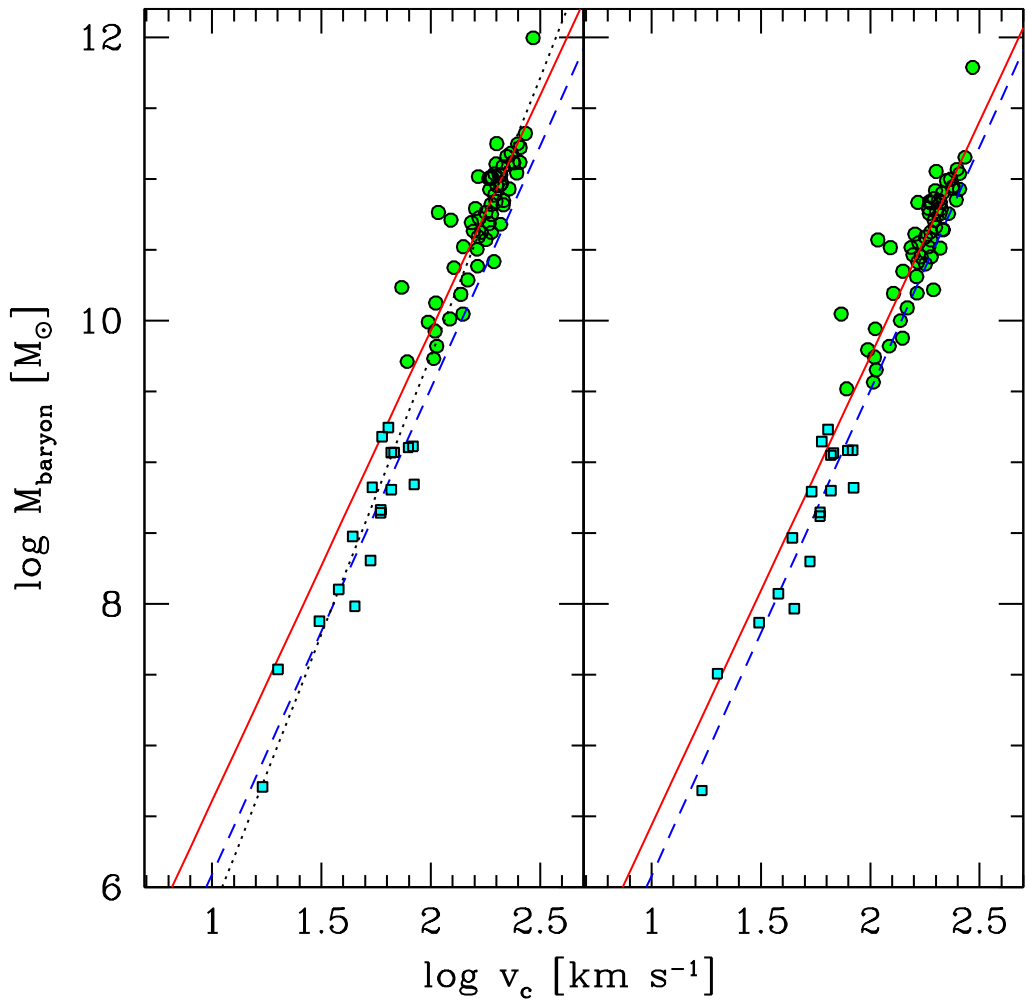}
\caption{Baryonic Tully Fisher for star dominated galaxies (green circles and red solid line) compared to that for gas dominated galaxies (cyan squares and blue dashed line). The data for gas dominated galaxies comes from \cite{mcgaugh12}. The left panel has our stellar masses renormalized to match on average those of \cite{mcgaugh12} using a subset of galaxies in common to the two samples. In the right panel we do not renormalize the stellar masses. The slopes of the fits done independently for the stellar and gaseous dominated samples are remarkably similar. The intercepts however vary. Because the intercepts vary less using our stellar mass estimates, we conclude that, under the assumption that a single, linear BTF is the correct underlying description of galaxies, our stellar mass estimates are to be preferred. The dotted line in the left panel, which is the bisector fit to the union of the two data sets results illustrates how fitting to the two samples together can result in a slope that is much larger than either sample suggests independently ($3.93 \pm 0.19$ in comparison to $3.32\pm 0.19$ (stellar) and $3.43 \pm 0.66$ (gaseous)).}
\label{fig:mcgaugh}
\end{figure*}

This is not a perfect test. The \cite{mcgaugh12} sample is treated differently than ours in a variety of ways. For example, the correction for metals is not applied (as these low mass galaxies have low metallicities) and resolved rotation curves are used to measure where the rotation curve is flat, $V_{f}$. Nevertheless, we can explore how each sample behaves independently as well as in combination. 
One factor we can correct for is the possibility of offsets in the stellar masses as calculated by \cite{mcgaugh12} and ourselves, to ensure that the two galaxy samples are on the same stellar mass system (even those that are gas dominated do have a stellar baryon component). For the eight galaxies in common to our samples, we find that on average $\log M_*$ differs by 0.215, with the \cite{mcgaugh12} masses being larger. We therefore ``correct" our values upward to provide a direct comparison to his results in the left panel of Figure \ref{fig:mcgaugh} and use these ``corrected" masses when applying the criteria to select gas- and star-dominated galaxies.  Using the same selected galaxies, but removing the ``correction" factor (and so reducing the quoted \cite{mcgaugh12} stellar masses), we obtain the results shown in the right panel of Figure \ref{fig:mcgaugh}. The two panels of the Figure therefore comprise a test of the stellar mass prescriptions under the assumption that a single BTF applies across the full galaxy $v_c$ range. 

There are various results of note in the Figure. First, the fits to either the gas- or star-dominated galaxies in either panel result in fits that have nearly the same slope but different intercepts. 
Using the ordinary least-squares bisector regression fitting algorithm \citep{isobe}, we find slopes of $3.32\pm0.19$ and $3.43\pm0.66$ for the stellar and gaseous samples when applying the correction to move our stellar mass estimates onto the \cite{mcgaugh12} scale. Without the correction, we find slopes of $3.31\pm0.14$ and $3.43\pm0.66$, respectively. It would appear that the change in stellar mass normalization makes no significant difference and that in both cases the slopes for the gas- and star dominated samples are similar. The lack of any effect on the slopes is a result of having selected gas and star dominated samples. In the former, changes in the stellar masses are nearly irrelevant and in the latter changes simply result in a zero-point shift of the stellar masses. This zero point change is evident in the result that between the corrected and uncorrected samples the intercept changes from 3.29 to 3.13. The similarity in slopes also argues that other differences between the samples, such as the use of $V_{f}$ vs. W$_{50}$ or treatment of the gas masses, also have little impact on the determination of the BTF slope. Nevertheless, an important next step would be to confirm this claim for a large sample of galaxies with both $V_f$ and W$_{50}$. Second, all of these results are consistent with the slope = 3.4 line plotted in Figure \ref{fig:bf}, suggesting good agreement with our theoretical expectations. Third, we  demonstrate the importance of the stellar mass normalization when fitting a single relationship to the combination of the gas and star dominated samples. In the left panel of Figure \ref{fig:mcgaugh} we also show this fit, which has slope $3.93\pm 019$. While this fit appears to be a good description of the data, we know that it reflects the slope of neither the gas or star dominated samples individually. In the case where we use our uncorrected stellar masses, the slope decreases to $3.68\pm0.17$ because the intercepts of our gas and star dominated galaxy fits are closer to each other. This result illustrates how one can obtain a much steeper BTF slope when combining gas and star dominated samples if the two have different intercepts (or mass normalizations).  We stress that this demonstration does not necessarily invalidate the \cite{mcgaugh12} results because we have not demonstrated such a failing within his self-consistent dataset, it only highlights the importance the stellar mass normalization can have on BTF slope determinations.

One could take the argument further and advocate a larger reduction of the stellar masses to improve the agreement between the gas and star dominated galaxy fits. Such a reduction would not be unwarranted because the \cite{eskew} relation is based on a Salpeter IMF. However, there are also uncertainties in the gas masses themselves, the sample is sufficiently sparse that the fitted parameters have large uncertainties, and the gas-dominated galaxies have not been measured in exactly the same manner as the S$^4$G sample. With larger samples of gas dominated galaxies that are observed in a consistent manner with how the star dominated ones are observed, it may be possible to obtain a more precise calibration of the stellar mass estimators, but such a treatment is currently premature.

Based on these results we conclude that our data favor a BTF slope between 3.3 and 3.7, and so quote $3.5 \pm 0.2$. This is in agreement with some previous determinations \citep{avila,hall} and that our stellar mass normalization is a potential explanation for why our slope is in disagreement with others determinations \citep[such as that of ][]{mcgaugh12}. 

These results are all predicated on a particular choice of initial mass function (IMF), and on its universality.
Recent results on early type galaxies \citep{treu,vandokkum,cappellari} and stellar clusters \citep{strader,zaritsky12,zaritsky13}, suggest that the IMF is not universal and that among the galaxies the IMF variations track total mass. If such a pattern exists also among disk galaxies, then it will affect the BTF slope. To gauge the magnitude of the effect, we adopt the relationship between $M/L$ and velocity dispersion found by \cite{cappellari}, adopt $v_c/\sqrt{2.5} \sim \sigma$ \citep{burstein, weiner,z06a,z12a}, and recalculate the BTF slope using the low scatter subsample. We find a magnitude change in slope  $<$ 0.1 (steeper), so subdominant to the current level of uncertainties arising from the cross calibration of the gas and star dominant galaxy samples. Eventually the BTF may also be a tool in addressing questions about IMF variations in disk galaxies, but greater precision is required.

\begin{figure*}
\plotone{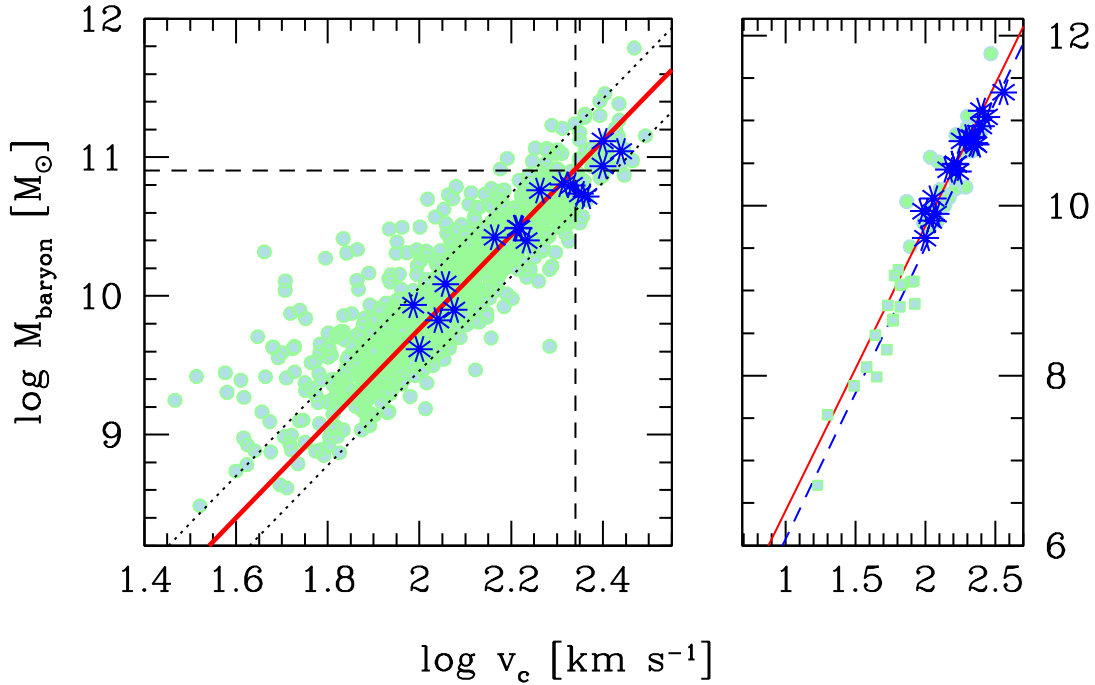}
\caption{Comparison of simulated galaxies from \cite{aumer} to our data. We reprise Figures \ref{fig:bf} and \ref{fig:mcgaugh} and superpose the simulated galaxies as blue stars. For the right panel we use our stellar masses (eg. the right panel of Figure \ref{fig:mcgaugh}). The excellent agreement between the simulations and the data demonstrates how well current simulations are doing at reproducing the mean trend. The scatter is somewhat less than observed, but the simulations aimed to reproduce dynamically quiet disk galaxies. The various lines are the same as in the previous respective Figures.}
\label{fig:aumer}
\end{figure*}

Finally, we consider the most recent simulations that address where disk galaxies fall in the $M_{baryon} - v_c$ space \citep{aumer}. Those simulations apply a multiphase SPH code with elaborate treatments of metal production, cooling rates, and metal diffusion to examine a range of galaxy properties. They include SN feedback, but not AGN feedback, and examine 16 simulated galaxies that range in halo mass from 10$^{11}$ to $3 \times 10^{12} M_\odot$. In comparing to the
\cite{mcgaugh12} results they found some discrepancies that they attributed to details of their adopted prescriptions. However, comparing to our data (Figure \ref{fig:aumer}), we find that their simulated galaxies do an excellent job of matching the properties of real galaxies. We conclude that current simulations are on track to reproduce the internal properties of galaxies, although the conflicting conclusions reached using either our or the \cite{mcgaugh12} sample can also be taken as a cautionary tale regarding the importance of unresolved uncertainties in the stellar mass determinations. 

We close by taking note of some tension between the BTF results, which indicate a constant $M_{baryon}/M_{halo}$ as a function of $M_{halo}$, and those obtained by comparing the statistical properties, such as number density and clustering, of galaxies to those of simulated dark matter halos \citep[see][and references therein for examples from a large set of pertinent references]{kauffmann,benson,white,berlind,yang,guo}. Specifically, all studies based on statistical properties of galaxies and simulated or measured dark matter halo masses, the empirical masses obtained via observations of gravitational lensing or satellite kinematics, indicate that the fraction of stellar mass, $M_{stellar}$, to dark matter mass, $M_{halo}$, peaks in galaxies with $M_{halo} \sim 10^{12} M_\odot$ and declines toward either more or less massive halos. TF and BTF have little to say on this issue at the large mass end, where rotationally supported galaxies are rare, but can inform results at the lower mass end. In fact, \cite{guo} note that their results, based on the statistical approach, are also consistent with TF, which due to the kink we noted above supports a lower $M_{stellar}/M_{halo}$ for galaxies with low $v_c$. The declining stellar mass fraction in lower mass halos mass could be the result of either a lower efficiency of star formation or the loss of baryonic matter from which to form stars in these halos. 
Recent work applying the statistical treatment but relying not only on optical luminosity measurements, which trace the stars, but also including H{\small I} measurements suggest that that low mass galaxies retain a lower fraction of their baryons \citep{papa}. This most recent result is in apparent conflict with the BTF results.

Both approaches purport to measure the baryon fraction and both have a history of consistent results that demonstrate their internal robustness. If we accept that each approach is free of unknown, and physically uninteresting, systematic errors, how can we resolve this conflict? One avenue might appeal to the fact that the results do not pertain to exactly the same systems, as the BTF results are for rotation-supported galaxies and the statistical ones are for the entire galaxy population and rotation supported galaxies become increasingly rarer at lower masses. Another approach might appeal to a higher degree of complexity in halo occupation models \citep[cf.][]{zhu,gao} even thought attempts to identify secondary factors beyond $M_{halo}$ have not identified such \citep[see][]{tinker}. Whatever the eventual resolution involves, the presence of even some galaxies with high $M_{baryon}/M_{halo}$ at low $M_{halo}$ demonstrates that the ability of a halo to retain its baryons is not solely related to $M_{halo}$, as is often envisaged in models that attempt to reproduce the decreasing $M_{baryon}/M_{halo}$ found by the statistical studies. Once again, rather than resolving the problem, we are only able to acknowledge the limits of our current understanding and urge caution in generalized treatments of galaxy formation and evolution, but also stress that simulations need to match both the statistical properties of galaxies and the individual global properties.

\section{Conclusions}

We examine the Tully-Fisher (TF) and Baryonic Tully-Fisher (BTF) using the new sample of IR photometry provided by the S$^4$G survey \citep{sheth,munoz}. The use of this large, homogenous sample brings the natural advantages of uniformity and large statistics, and also provides photometry that is less susceptible to reddening and stellar population variations. We apply a new empirical stellar mass estimator from \cite{eskew} to obtain stellar masses for the sample. When combined with the available large, homogenized archival H{\small I} data of the Cosmic Flows program and complemented with additional observations \citep{courtois11}, we have a primary sample of 903 galaxies for study.

We confirm the superiority of the BTF for lower mass galaxies and that the overall scaling relation becomes noticeably more linear than the TF. Furthermore, we quantify the scatter introduced by various  lax selection criteria related to the inclination limits of the sample, the morphological type range of the sample, and the effect of peculiar velocities on nearby galaxies. In the end, we are able to identify a BTF with a scatter of 0.18 dex, without detailed pruning of galaxies with either morphological or kinematic anomalies. This is a fair sample to compare to simulations where such cuts are difficult to reproduce.

We demonstrate that physical scale, here parameterized by the half light radius, correlates with the residual in the BTF. However, the gain in precision is exceedingly slight and the standard BTF ($\log M_{baryon}$ vs. $\log v_c$) is to be preferred for this galaxy sample for its simplicity. 

We measure a BTF slope of $3.5 \pm 0.2$, consistent with the expectation where galaxies concentrate a fixed fraction of their initial baryons in the central, detectable components (stars and gas). Using the MW to normalize the relation, we show that the expectation of this simple model, where 0.07 of the total mass of the halo (or alternatively about 40\% of the baryons) is ``condensed" onto the central regions of the halo, is an excellent fit to the data. This agreement demonstrates that, independent of circular velocities for the range of velocities explored here, the resulting effects of physical processes that could have affected the global properties of these galaxies (outflows, inflows) are independent of halo mass. Our sample also supports galaxy-to-galaxy scatter in the condensed baryon fraction up to a factor of two. Untangling the observational and intrinsic scatter is difficult with the current data, but manifestly a next avenue that needs to be explored.

Our value of the BTF slope agrees with certain previous studies \citep{avila, hall} and we present one possible explanation for the disagreement with other studies that found steeper slopes \citep{mcgaugh12}. We suggest that the disagreement comes about from the normalization of the stellar mass estimation. While we do not have direct evidence favoring one mass normalization over another, we prefer ours because it results in closer agreement in the derived BTF relations when using independent samples of gas and star dominated galaxies. We show that combining the gas and star dominated galaxy samples, when different stellar mass normalizations are used, can result in slope differences consistent with what is found among the various studies. Our data also result in excellent agreement with the recent simulations of \cite{aumer}. Although that agreement cannot be used to support our stellar mass estimates, it does suggest that simulations are approaching a level of sophistication rivaling the observational uncertainties in determining the internal structural properties of individual galaxies. Finally, and perhaps eventually most illuminating when the origin is understood, we note the continuing tension between the BTF results, which suggests a constant $M_{baryon}/M_{halo}$, and those from statistical studies of galaxies, such as abundance matching models, that find a decreasing $M_{baryon}/M_{halo}$ with decreasing $M_{halo}$.

\begin{acknowledgments}

DZ acknowledges financial support from 
NASA ADAP NNX12AE27G and NSF AST-1311326, and thanks NYU CCPP for its hospitality during long-term visits. The authors thank Michael Aumer for sending us the baryonic masses and circular velocities of simulated galaxies that we incorporated into Figure \ref{fig:aumer}, and Marc Verheijen and R. Brent Tully for comments on the manuscript. We also gratefully acknowledge discussions with Stacy McGaugh, which led both to improvements of the current manuscript and to related ideas to pursue in subsequent work.
HC and JS acknowledge support from the Lyon Institute of Origins under grant ANR-10-LABX-66.
The authors acknowledge the support from the FP7 Marie Curie Actions of the European Commission, via the Initial Training Network DAGAL under REA grant agreement PITN-GA-2011-289313.
The authors thank the entire  S$^4$G team for the efforts in making this program possible. 
K.S., J-C.M-M, and T.K acknowledge support from the National Radio Astronomy Observatory is a facility of the National Science Foundation operated under cooperative agreement by Associated Universities, Inc.
This research has made use of the NASA/IPAC Extragalactic Database (NED), which is operated by the Jet Propulsion Laboratory, California Institute of Technology, under contract with NASA. This research 
is based in part on observations made with the Spitzer Space Telescope, which is operated by the Jet Propulsion Laboratory, California Institute of Technology under a contract with NASA.
\end{acknowledgments}

\end{document}